\documentclass[12pt, oneside]{article}   	
\usepackage{csquotes}
\usepackage{comment}
\usepackage{subfigure}
\usepackage{xcolor}
\usepackage{booktabs}
\usepackage{array}
\usepackage{url}
\newcolumntype{P}[1]{>{\raggedright\arraybackslash}p{#1}}
\usepackage{authblk}
\usepackage{setspace}
\usepackage{geometry}
\usepackage{graphicx}
\usepackage{amssymb}
\usepackage{amsmath}
\usepackage{adjustbox}
\usepackage{breakcites}
\usepackage{footnote}
\usepackage{natbib}

\usepackage{breakurl}
\usepackage{hyperref}
\title{Methodological Problems in Every Black-Box Study of Forensic Firearm Comparisons}

\author[a,b,f]{Maria Cuellar}
\author[d,f]{Susan Vanderplas}
\author[c,f]{Amanda Luby}
\author[e]{Michael Rosenblum}

\affil[a]{Department of Criminology, University of Pennsylvania, 3718 Locust Walk, Philadelphia, PA, 19104, United States}

\affil[b]{Department of Statistics and Data Science, Wharton School, University of Pennsylvania, Walnut Street, Philadelphia, PA 19104, United States}

\affil[c]{Department of Mathematics and Statistics, Swarthmore College, 500 College Avenue, Swarthmore, PA 19081}

\affil[d]{Department of Statistics, University of Nebraska-Lincoln, 340 Hardin Hall North Wing Lincoln, NE 68583-0963}

\affil[e]{Department of Biostatistics, Johns Hopkins University, Bloomberg School of Public Health, 615 N. Wolfe Street, Baltimore, MD 21205}

\affil[f]{Center for Statistics and Applications in Forensics Evidence (CSAFE), Iowa State University, 613 Morrill Road, Ames, IA, 50011, United States}

\begin{document}
\maketitle

\newpage

\begin{abstract}
Reviews conducted by the National Academy of Sciences (2009) and the President's Council of Advisors on Science and Technology (2016) concluded that the field of forensic firearm comparisons has not been demonstrated to be scientifically valid. Scientific validity requires adequately designed studies of firearm examiner performance in terms of accuracy, repeatability, and reproducibility. Researchers have performed ``black-box'' studies with the goal of estimating these performance measures. As statisticians with expertise in experimental design, we conducted a literature search of such studies to date and then evaluated the design and statistical analysis methods used in each study. Our conclusion is that all studies in our literature search have methodological flaws that are so grave that they render the studies invalid, that is, incapable of establishing scientific validity of the field of firearms examination. Notably, error rates among firearms examiners, both collectively and individually, remain unknown. Therefore, statements about the common origin of bullets or cartridge cases that are based on examination of ``individual" characteristics do not have a scientific basis. 
We provide some recommendations for the design and analysis of future studies.
\end{abstract}

\section{Introduction} \label{sec:intro}

The National Academy of Sciences Report ``Strengthening Forensic Science in the United States: A Path Forward" \citep{NAS2009} expressed a critical need for research aimed at establishing the scientific foundations of forensic science. Among other conclusions about firearms examination, it stated that ``sufficient studies have not been done to understand the reliability and repeatability of the methods.'' \citep[p.154]{NAS2009}. The President's Council of Advisors on Science and Technology \citep{PCAST} evaluated the scientific validity of some forensic methods, including firearms examination, that are used in criminal courts. They similarly concluded that the scientific validity of firearms examination has not been established.

\cite{PCAST} explained why empirical studies to evaluate examiner performance, called validation studies, are of critical importance: ``Without appropriate estimates of accuracy, an examiner's statement that two samples are similar--or even indistinguishable--is scientifically meaningless: it has no probative value, and considerable potential for prejudicial impact. Nothing--not training, personal experience nor professional practices--can substitute for adequate empirical demonstration of accuracy.'' \citep[p.46]{PCAST}. We agree, since a fundamental principle of sound science is the need for demonstration by repeated experiments \cite[p.38]{NASSoundScience}. The importance of empirical studies is also highlighted by the U.S. National Institute of Standards and Technology (NIST) in their report on evaluating scientific foundations of forensic methods \citep{Butleretal2020_NIST_Foundation_Reviews}. This motivates our focus on validation studies. 

In addition to the aforementioned NAS and PCAST reports, there is a substantial literature describing problems with the study design and statistical analysis methods in validation studies of firearms examination, e.g., 
\citet{spiegelman2012analysis, dror2020misuse, hofmann2020treatment, dorfman2022inconclusives, dorfman2022reanalysis, scurich2022inconclusives, khan2023hierarchical, khan2023shining,doi:10.1073/pnas.2301843120,10.1093/lpr/mgad010}, among others. 
Our contribution builds on this prior work. We conducted a literature search of validation studies (focusing on black-box studies) of firearms examination to date, and then evaluated the design and statistical analysis methods used in each. We identified serious methodological flaws that are common to all such studies, with minor exceptions (see Table~\ref{tab:studydesignflaws} of Section~\ref{sec:summary_of_key_flaws}). Most of these flaws were presented in the aforementioned prior work for a subset of studies; we show that these flaws are not the exception, but instead are generally the rule across such studies. Our search also revealed new flaws. We present this critique constructively with the goal of raising awareness of how to improve future studies.

In the next section, we explain why sound  experimental design and statistical analysis methodology used in validation studies of medical diagnostic tests are  applicable to validation studies of firearms examination. In Section~\ref{sec:blackboxstudies} we describe our literature search of  firearms examination validation studies. Study design and analysis flaws are summarized in Section~\ref{sec:summary_of_key_flaws} and explained in detail in Section~\ref{sec:flaw_details}. We end with a discussion in Section~\ref{sec:discussion}.

\section{Relevance of research and evaluation methods used in medical diagnostic testing} \label{sec:relevanceDiagnosticTesting}

The 2009 National Academies of Sciences report \citep{NAS2009} summarized the need for scientifically sound evaluations of forensic analysis methods, including firearms examination. It highlighted the relevance of the sound research and evaluation methods used in medical diagnostic testing, in the following passage: 

\begin{displayquote}
 A body of research is required to establish the limits and measures of performance and to address the impact of sources of variability and potential bias. Such research is sorely needed, but it seems to be lacking in most of the forensic disciplines that rely on subjective assessments of matching characteristics. These disciplines need to develop rigorous protocols to guide these subjective interpretations and pursue equally rigorous research and evaluation programs. \begin{bf}The development of such research programs can benefit significantly from other areas,  notably from the large body of research on the evaluation of observer performance in diagnostic medicine\dots  \end{bf} \citep[p.8]{NAS2009}
\end{displayquote}

We bolded the last sentence to highlight the relevance of medical diagnostic tests. There are many similarities between diagnostic testing involving human observers, e.g., breast cancer diagnostic testing using mammography, and firearms examination. In both mammography and firearms examination, a human observer examines images and based on their experience and training makes a subjective decision about it. In mammography an examiner decides whether the images (taken from different angles and possibly magnified) indicate cancer or not, while in firearms examination an examiner decides whether bullets/cartridge cases (using magnified images) were fired from the same or different guns. Both involve subjective analysis followed by a yes/no decision (with a possibility for inconclusive results). 

Firearms examiners have drawn parallels between diagnostic testing and their field, e.g., \citet[p.203]{bunch2003comprehensive}, state that ``For data analysis purposes, a forensic comparison examination can be considered analogous to a clinical test such as a blood test.''  We therefore refer to published guidance documents from the U.S. Food and Drug Administration  and the Clinical and Laboratory Standards Institute on how to design and analyze validation studies. Their recommendations are applicable to any diagnostic test procedure, including ones that involve human observation and subjective decisions, and therefore can be applied to firearms examination.

\section{Selection of black-box studies} \label{sec:blackboxstudies}

Firearms examination is a subjective method, i.e., key parts of the comparison rely on human judgment. Because of this, validation studies need to be black-box, i.e., they ask firearm examiners to make decisions about a series of test cases where the experimenter (but not the firearm examiners) knows the ground truth, and performance is measured by comparing firearm examiners' decisions to the ground truth across test cases. These are called ``black-box'' because they measure whether decisions are correct or not, while treating the examiner's internal decision process (which is difficult to quantify due to its subjective nature) as a ``black-box". We use the term ``black-box studies" to refer to black-box validation studies. 

We compiled our list of black-box studies of firearms examination (see list in Appendix A) from multiple sources. 
Our criteria for including a black-box study in our list in Appendix A are the following: (i) it includes human examiners solving test cases involving fired cartridge cases and/or bullets with known ground truth; (ii) it is not a pilot/preliminary/exploratory study. 

We first searched for studies by using specific terms (i.e., ``validation", ``firearms", ``examination", and ``comparison") in the Association of Firearms and Toolmarks Examiners (AFTE) Journal Article Index. We then considered studies from the Interpol review papers covering the period 2012-2022; that publication compiles  validation studies of firearms examination (among other forensic methods) from the following journals: Australian Journal of Forensic Sciences, Forensic Science International, Forensic Science International: Synergy, Forensic Sciences Research, Journal of Forensic Identification, Journal of Forensic Sciences, Science and Justice. 

We also considered the studies listed in the \cite{PCAST} report and in a response\footnote{This response was not endorsed by OSAC nor NIST, but in an effort to be as inclusive of potential black-box studies as possible, we considered the studies that it listed.} to this report by the 
\cite{OSAC_SWGGUN_response_to_PCAST}. We conferred with a practicing firearms examiner to ask for additional studies. Each author of this manuscript included any studies that they are familiar with through previous work on the scientific validity of firearms examination. 

A total of 28 different studies meeting our inclusion criteria were obtained. These are listed in Appendix A. (There are 33 citations there, but some refer to the same study.)

\section{Evaluation of study designs and statistical analyses} \label{sec:summary_of_key_flaws}
A validation study needs to have sound experimental design and statistical analysis principles to provide reliable results. Some standard principles include the following: selecting participants and materials that are representative of the full spectrum of real casework; including sufficient numbers of such participants and materials to answer the scientific question with the desired level of precision; estimating error rates correctly along with valid confidence intervals; and addressing missing data with appropriate statistical methods.  With minor exceptions described below, none of the 28 studies adhered to any of these key principles. Although it can be challenging for validation studies of firearms examination to adhere to all of these principles, it is possible and is necessary in order to produce reliable results. The challenge of adhering to all of these principles is  faced in other domains, e.g., in medical-imaging validation studies, where these challenges have been successfully met.

Table 1 lists study design and analysis flaws that we found in every study in Appendix A, with some exceptions for flaws C and E indicated by footnotes. It is not an exhaustive list. Each of the methodological flaws B-E in Table~\ref{tab:studydesignflaws} is so consequential that having even one such flaw renders a validation study scientifically  unsound. 
Also, flaw A is important because it exacerbates the other flaws, as described below. 
Most of the flaws in Table~\ref{tab:studydesignflaws} have been identified for at least some studies in prior work, as described in Section~\ref{sec:intro}. Our contribution is to summarize the  flaws that are both common and consequential across the black-box studies in Appendix A. Because these flaws are shared across studies, it is not possible to combine results from the studies to overcome the flaws. The only way to demonstrate scientific validity is to conduct future, adequately designed and analyzed studies.

\begin{savenotes}
\begin{table}[ht]
    \centering
    \normalsize
    \begin{tabular}{@{}P{0.7\textwidth}P{0.25\textwidth}@{}}
      \textbf{Flaw} & \textbf{Impact} \\
      \midrule
      \textbf{A. Inadequate sample size:} No sample size calculation was done to determine how many firearms, examiners, and bullets/cartridge cases are needed to achieve the study goals. The sample size calculation is ``one of the most important parts of any experimental design problem." \citep[p.44]{montgomery2017design}. & 
        Insufficient number of firearms are used, leading to low precision. \\
        \textbf{B. Non--representative sample:} Study conditions, materials (firearms and ammunition), and participants (examiners) are not representative of the full spectrum of real casework.  &
        Results not applicable to real casework. \\
        \textbf{C. Error rates incorrectly computed}  because inconclusive responses are essentially treated as correct or ignored.\footnote{One study had no inconclusive responses and some studies did not report a study-wide error rate. Some studies computed error rates in multiple ways, but all are flawed.}  & 
        Error rates may be underestimated.\\
        \textbf{D. Invalid confidence intervals for error rates}. Confidence intervals do not account for the structure of study designs (multiple examiners and multiple firearms, which leads to statistical dependence), or no confidence interval is provided.& 
        Reported range of plausible error rates in studies is too small or is unknown \\
        \textbf{E. Missing data.}\footnote{Some of the smaller studies had no missing data.} Missing data, such as item and unit non--response, are not handled with appropriate statistical methods. & 
        Error rates may be biased. \\
        \bottomrule
    \end{tabular}
    \caption{Flaws with the study design and statistical analysis that occur in every black-box study that we analyzed, along with the impact of each flaw.}
    \label{tab:studydesignflaws}
\end{table}
\end{savenotes}

Prior work has identified some of the flaws.
For example, \cite{spiegelman2012analysis} found flaws A, B, D, and E, as well as other flaws not on our list, in several firearms examination studies.
\citet{hofmann2020treatment} study the treatment of inconclusive responses in error rate calculations (flaw C). 
\citet{khan2023hierarchical,khan2023shining} consider some large validation studies and show that they have high rates of ``missingness'', i.e., unit and item nonresponse (flaw E). 

\section{Flaws with study design and statistical analysis} \label{sec:flaw_details}

\subsection{Flaw A: Sample size} \label{sec:samplesizeflawdetails}

Contrary to standard practice in the experimental design of validation studies, none of the black-box studies in Appendix A carried out the important step of computing the sample size, i.e., the number of examiners, the number of firearms of each type, the number of ammunition types, and the number of bullets/cartridge cases required to achieve the study goals. According to the textbook {\em Design and Analysis of Experiments} by \citet[p.44]{montgomery2017design}, ``Selection of an appropriate sample size is one of the most important parts of any experimental design problem.''   This requirement is stated in FDA guidelines for validating diagnostic tests \citep[p.42]{FDA2007statistical}, which states that  ``The study should be of sufficient size to evaluate the candidate test based on the desired level of uncertainty.'' The sample size needs to  be  determined prior to starting the study and is computed in order to provide a high likelihood of achieving the study's goals. Without a sample size calculation, there is no reason to expect that a validation study will be able to do this. 

The \citet[p.13]{OSAC2020humanfactors} also acknowledges the need for a sample size calculation when it states that, ``To properly assess the accuracy of a method at any difficulty level, it is important to include adequate numbers of test specimens at that level.'' It also recommends that, ``Statisticians or other experts familiar with statistical power and sample size requirements for experimental research can assist in determining the appropriate numbers of test specimens and examiners for validation and reliability studies.'' As statisticians, we agree with these statements. Skipping this important step can lead to studies that fail to estimate accuracy precisely. This occurred in all of the studies in Appendix A. Not doing a sample size calculation may be acceptable in pilot studies or exploratory studies, but not in validation studies. 

\citet[p.262]{kerkhoff2018part} state that, for their firearms examination study, ``The sample size was relatively small (53 uncorrected, 39 corrected conclusions) and a much larger sample would be needed to get a good estimate of the rate of misleading evidence in practice.'' We applaud \citet{kerkhoff2018part} for clearly stating the limitations of their study and for stressing the need for adequate sample size and proper study design more generally. The quote above explains why we did not include the studies of \cite{Kerkhoff2015DesignAR,kerkhoff2018part,neuman2022blind} in our list of black-box studies, i.e., they are exploratory studies. Such studies are valuable for hypothesis generation and for informing the planning of future studies; however, they are not appropriate for validation.

The sample size calculation depends on what type of experimental design is used. Almost every black-box study in Appendix A uses a design where one or more test cases from each firearm (or firearm pair) is separately evaluated by multiple examiners, and similarly each examiner evaluates multiple test cases. In such designs, a correct sample size calculation needs to simultaneously account for (i) dependence among test cases evaluated by the same examiner and (ii) dependence among test cases involving the same  firearm(s) \citep{FDA2007statistical,montgomery2017design}. This general setup is called a multi-reader, multi-case (MRMC) design, and there is a rich statistical literature on how to appropriately analyze it, e.g., \cite{zhou2009statistical,gallas2009framework}. 

\citet[pp.219--228]{zhou2009statistical} provides detailed guidance about how to calculate sample sizes in MRMC studies, while appropriately handling dependencies that arise from the experimental design. The software package iMRMC in R by \cite{imrmc} does  sample size calculations, is free to download, and is provided by the FDA. This tool can be used to determine how many firearms, ammunition types, examiners, and test cases are needed.\footnote{If the estimated false positive rate is close to 0 or 1, one may need to use exact methods to construct confidence intervals, e.g., generalizing the Clopper-Pearson method described in Appendix B.}

\subsection{Flaw B: Representative samples} \label{sec_representative_samples}

For a set of studies to assess the accuracy, repeatability, and replicability of a pattern-comparison method such as firearms comparison, it is critical to include representative samples, i.e., a set of participants, materials, and test conditions that are representative of the full spectrum encountered in real casework.
As we explain below, none of the studies that we reviewed gives any evidence that its conditions, participants, and materials (firearms and ammunition) are representative of the full spectrum of real casework. But first we consider statements by several scientific groups about the importance of having a representative sample.

The American Statistical Association (ASA) is the largest community of statisticians in the world and includes members from over 90 countries working in government, academia, and industry; their goal is ``promoting sound statistical practice" \citep{asawebsite}. According to the {\em ASA Position on Statistical Statements for Forensic Evidence}, ``To be applicable to casework, rigorous empirical studies of the reliability and accuracy of forensic science practitioners' judgments must involve materials and comparisons that are representative of evidence from casework'' \citep{asastatement}. Thus, the ASA also agrees that representation is important.

We next consider the report {\em NIST Scientific Foundation Reviews} by the the National Institute of Standards and Technology 
\cite[p.2]{Butleretal2020_NIST_Foundation_Reviews}. The report describes key principles and criteria for evaluating studies of the scientific validity of forensic methods. One of the key criteria is the following: ``Are test results fit for purpose (i.e., are they expected to reflect performance under casework conditions)?" In other words, a representative sample is needed. The \citet[p.6]{nifs2019empirical} also provides a checklist that includes this point.

 \citet[pp.7-8]{monson2022planning} state that ``To detect a small or rare effect (examiner errors in the present case) one must make a large number of observations within a representative population.'' Though we disagree that examiner error rates have been demonstrated to be small or rare, we do agree that in order to have any opportunity to demonstrate this, one needs a large number of observations from a representative population.  This is especially important since some firearm types produce harder/easier to classify toolmarks, and so assessing a relatively small number of firearm types may lead to conclusions that don't generalize to what's commonly encountered in real casework.

The FDA warns about the dangers of not including a representative sample. The ``FDA recommends the set of subjects and specimens to be tested include: subjects/specimens across the entire range of disease states\dots'' and that ``If the set of subjects and specimens to be evaluated in the study is not sufficiently representative of the intended use population, the estimates of diagnostic accuracy can be biased.'' \cite[p.18]{FDA2007statistical}.

A prerequisite to conducting experiments to evaluate reliability of firearms examination is to first characterize the full spectrum of firearms (and firearms manufacturing techniques) and ammunition known to be used in crimes, as best possible given available data sources, or new data sets should be created for this purpose. Only after this is done can studies be designed that include representative samples of firearm types. An analogous issue arises for ammunition and for examiners, e.g., with respect to level of training and to protocols used in different crime labs. 

  \cite{khan2023shining} consider how representative the sample of firearm examiners who participated in the study of \cite{monson2023accuracy} are with respect to 
characteristics that may affect error rates, such as years of expertise, laboratory protocols, and amount of training. They show that this sample of examiners is not representative of the population of examiners, which may lead to underestimating error rates. 

It is not necessary to include every possible firearm, ammunition type, or examiner. What is needed is a representative sample, which can be obtained by first characterizing the full spectrum of firearms, ammunition, and examiners and then taking a random or systematic sample from these. This is not easy, but it is a necessary step since otherwise it is not possible to determine whether a sample is representative of the full spectrum of real casework.

Study conditions should also be representative of real casework, so that study results generalize to real casework conditions. A discrepancy between study and casework conditions is that  examiners in a study know that their answers will be compared to known ground truth. This creates a potential motivation to modify behavior, e.g.,
they may spend more time analyzing the samples and/or may be less likely to make an identification decision. 
Changes in behavior could lead to error rates in studies that are not the same as the error rates in real casework \citep{dror2020misuse}. One way to partially address this is to conduct studies that insert  test cases with known ground truth into the workflow of real casework in a blinded manner (i.e., without informing examiners which samples are test cases or not). This has been pilot tested in quality control programs  \citep{kerkhoff2018part,neuman2022blind}. \cite{mejia2020implementing} consider the logistical challenges in implementing such blinded programs and provide recommendations on how to overcome them.

Another discrepancy between conditions in studies and real casework is the presence of contextual information in the latter that may bias decision making \citep{KASSIN201342,MATTIJSSEN2016113}. Contextual information could include, for example, other evidence in a case. According to the aforementioned related work, such contextual bias may be reduced in real casework if examiners are blinded to all information about a case other than what is needed to compare bullets/cartridge cases.

Additional factors have potential to impact the difficulty of classifying toolmarks; see e.g., the list of factors in \citep[Section 1.2]{spiegelman2012analysis}. For example, the condition of firearms (e.g., newly manufactured versus damaged/corroded) may impact the difficulty of classifying toolmarks, yet this has not been systematically varied and analyzed in any black-box study, to the best of our knowledge. None of the black-box studies in the Appendix reports the use of damaged/corroded firearms for generating test cases.  Therefore, it’s unknown whether the error rates from these studies would apply to such firearms.

\subsection{Flaw C: Treatment of inconclusives}
In firearms comparisons, examiners assess whether a pair of samples (bullets/cartridge cases) were fired from the same firearm. 
Roughly speaking, they select one of three conclusions for each pair: identification, elimination, or inconclusive.\footnote{Different subcategories are also possible, as well as being ``unsuitable" for evaluation; also, some laboratories report likelihood ratios instead of categorical responses. We focus on the three aforementioned conclusions for clarity of explanation.} A black-box study of sample comparisons should report the false positive rate, the false negative rate, the true positive rate, the true negative rate -- alternatively, the sensitivity and specificity. It is not sufficient to only report the false positive rate, for example. In the extreme case, all the participants in a study could mark all pairs as eliminations -- this would result in a zero percent false-positive rate, but it would be trivially uninformative. In addition, the study should report the treatment  of inconclusives. 

In most black-box studies that we reviewed, the ``inconclusive'' responses are effectively treated as correct or ignored when computing error rates\footnote{Some studies included additional ways to treat inconclusives. \cite{lawmorris2021}  present results treating inconclusives as correct, incorrect, and using a consensus-based scoring approach; they do not report a study-wide error rate. Some studies either did not report error rates but rather number of decisions \citep{Hambyetal2009}, while other studies resulted in no inconclusives \citep{Fadul_2011,Cazes_Goudeau_2013}. \cite{mattijssen2020validity, Mattijssen2021} used a `forced choice' design and reported error rates both including and excluding inconclusives. None of the aforementioned methods, however, adequately handles how to account for inconclusive responses when computing study-wide error rates.}. Both approaches (treating ``inconclusives'' as correct or ignoring them) are invalid and can lead to underestimation of error rates. The first approach artificially reduces the error rates because inconclusive responses get counted in the denominator but not the numerator when computing each error rate. For both approaches, there may be  an incentive for examiners to opt out of the most challenging test cases by responding ``inconclusive'', knowing that this will either decrease or have no impact on the study's error rate.

\citet{hofmann2020treatment} provide an in-depth exploration about inconclusives and their different treatments in black-box studies. They find that the rates of inconclusives may vary depending on the norms for training and reporting in different regions. They also find evidence that the inconclusive rate is higher for different-source than same-source samples. 

In the context of evaluating diagnostic tests,  a guidance from the \citet[p.20]{FDA2007statistical}  lists four statistical practices that it calls ``inappropriate''. The second ``inappropriate'' practice on this list is to ``discard equivocal new test results when calculating measures of diagnostic accuracy or agreement''.  In the context of evaluating the accuracy of firearms examination, this means that ``inconclusives'' should not be ignored nor omitted when computing error rates, as is done in some of the studies in Appendix A. 

The aforementioned FDA guidance also has an entire section on how to handle ``inconclusive'' results (also called ``equivocal'' or ``indeterminate'' results) titled: ``Avoid elimination of equivocal results'', that recommends not to ignore inconclusives \citep[p.18]{FDA2007statistical}. There, the FDA describes a procedure to handle ``equivocal'' results, which is to report error rates in the following two ways: first, set all ``equivocal'' responses to ``positive'' and compute error rates; second, set them all to ``negative'' and recompute error rates. 

The corresponding procedure in the context of firearms examination would be to set all ``inconclusive'' responses to ``identification'' and compute error rates; second, set them all to ``elimination'' and recompute error rates. We recommend this procedure since it gives estimated lower and upper bounds on the error rates. 
Applying this procedure to the cartridge case data from \citet{monson2023accuracy}, the corresponding false positive error rates are 51.4\% and 0.92\%, respectively; the corresponding false negative error rates are 1.76\% and 25.6\%, respectively. This is for illustration only, temporarily setting aside the other flaws in Table 1. All such flaws need to be addressed by future studies in order to produce valid results.

Another measure of accuracy is the likelihood ratio, which is discussed and computed by \cite{guyll2023validity}. Computing a study-wide likelihood ratio could be a useful approach, as discussed by \cite{10.1093/lpr/mgi008,thompson2013role,doi:https://doi.org/10.1002/9781118492475.ch4,champod2016enfsi, lund2017likelihood,ommen2018building}. However, because the study of \cite{guyll2023validity} has flaws A, B, D, and E, the reported likelihood ratios are invalid. For additional methodological flaws in that study, please see \citep{10.1093/lpr/mgad010}.

Another related issue is conformance, i.e., the need for examiners in black-box studies to  adhere to a clearly defined protocol. A lack of conformance has been described by \cite{BALDWIN2023111739}, who show that the meaning of inconclusive responses varies by  examiner. They explain that forensic laboratories vary in whether they  allow elimination decisions to be based on ``individual characteristics". Labs that do not allow this are in conflict with how the AFTE Range of Conclusions \citep{AFTErange} defines inconclusive and elimination responses. We put ``individual characteristics" in quotes because it is unknown whether these characteristics (surface contour patterns) constitute a signature that is unique to each individual firearm, up to the level that can be observed using current technology such as comparison microscopes, 3D imaging, etc. \citep{nas_ballistic_imaging_2008}. The \cite{AFTEposition} Theory of Identification  assumes the truth of this unproven uniqueness.

\subsection{Flaw D: Confidence intervals for error rates}
\label{sec:confidenceIntervals}
We first describe what confidence intervals are and why they are critical for interpreting the results of firearms examination studies. Next, we explain why the statistical analysis methods used to compute confidence intervals in all the black-box firearms examination studies in Appendix A are invalid.

Each of the black-box studies reports estimates (also called ``point estimates'') of error rates. For example, the false positive error rate is typically estimated by dividing the number of false identifications by the total number of different-source comparisons. For example, the point estimate of the false positive error rate for cartridge cases in the \citet{monson2023accuracy} study is 0.933\%. This point estimate alone is insufficient to draw any statistical conclusions (called statistical inferences) about the false positive error rate for firearms examiners (even when putting aside all of the other study design and analysis flaws in Table 1). The reason is that we also need to have some measure of uncertainty such as a confidence interval to put the point estimate into perspective. A confidence interval for the false positive error rate represents the range of plausible values for it that are consistent with the study data. The importance of reporting measures of uncertainty in addition to point estimates is emphasized in FDA recognized clinical guidelines for evaluating diagnostic tests \citep[p.42]{CLSI2023} which states that ``Point estimates alone are not enough to evaluate test performance because they do not reflect the variability in the estimates.''  

The National Academies of Science \citep[p.116]{NAS2009} also highlights the importance of providing confidence intervals that reflect the sources of variability. They state the following: 

\begin{displayquote}
A key task for the scientific investigator designing and conducting a scientific study, as well as for the analyst applying a scientific method to conduct a particular analysis, is to identify as many sources of error as possible, to control or to eliminate as many as possible, and to estimate the magnitude of remaining errors so that the conclusions drawn from the study are valid. Numerical data reported in a scientific paper include not just a single value (point estimate) but also a range of plausible values (e.g., a confidence interval, or interval of uncertainty).
\end{displayquote}

Confidence intervals are closely related to the margin of error in an opinion poll. For instance, consider an opinion poll asking likely voters whether they will vote for candidate A or B. As a simple example, if the opinion poll reports an estimate that 48\% will vote for candidate A and 52\% for candidate B, with a 1\% margin of error, then candidate B is in good shape. However, the same point estimates with a 5\% margin of error means that neither candidate is clearly in the lead since the uncertainty (as measured by the margin of error) is too high. Similarly, a confidence interval represents the uncertainty, i.e., the range of plausible values for the quantity of interest (e.g., the false positive error rate) that are consistent with the data. The margin of error is related to the sample size of a study in that a smaller sample size leads to a larger margin of error, ceteris paribus.

Though different methods are used to construct confidence intervals in each of the black-box studies, they all suffer from (at least) the same major flaw. The flaw is that none of them accounts for variation among firearms in their likelihood of producing easier/harder to match toolmarks. This variation can arise from different makes/models of firearms (such as a 9mm Glock vs, Ruger SR9) or from individual firearms of the same make/model such as the 27 Beretta M9A3-FDE semiautomatic pistols used by  \cite{amesii,monson2023accuracy}. The problem is that failing to account for such variation can (and did) lead to invalid confidence intervals, as described below.

Some of the largest studies, e.g., \cite{guyll2023validity} and the study of \cite{amesii} (which we refer to as ``AMES II"), show differences between firearm makes/models (such as Jimenez vs. Beretta) in their likelihood of producing easier/harder to match toolmarks. There is also evidence for the existence of such variation among individual firearms of the same make/model from the AMES II study. For example, examiners were asked to rate the difficulty of test cases as being “easy”, “average”, or “hard”.\footnote{A limitation is that the data here are examiners’ subjective assessments of difficulty and do not necessarily reflect actual difficulty.}  For each of the 27 Beretta pistols the ratings were averaged across all examiners and reported in Table F1 of the AMES II study \citep[p.113]{amesii}. For cartridge case comparisons, the proportion of “hard” test cases generated by a Beretta pistol ranged from 0\% to 62\% (for Beretta pistols “F” and “E”, respectively) when the ground truth was same source. The range was 1\% to 43\% (for Beretta pistols “O” and “Z”, respectively) when the ground truth was different source. This shows substantial variability among individual Beretta pistols (all of the same model) in terms of how difficult the resulting test cases were judged to be by examiners. As another example, the proportion of examiner evaluations that were “exclusions” varied widely, i.e., from 26\% to 86\%, depending on which of the 27 Beretta pistols was used to generate cartridge case comparisons when the ground truth was different source \citep[p.116, Table F3]{amesii}.  The analogous result for “identifications” under same source ground truth is a range from 60\% to 97\%.

Next, consider the black-box study of \citep[p.12, Appendix A]{mattijssen2020validity}\footnote{A caveat is that test cases in this study are, according to the authors, intended to be challenging and examiners make decisions based only on images of firing pin aperture shear marks.}  which used 38 9mm Luger Glock pistols to generate same-source test cases. Each pistol was fired twice and each of the resulting 38 pairs of cartridge cases was evaluated by 77 examiners. For each of the 38 cartridge case pairs, the true positive rate averaged over 77 examiners was estimated by dividing the number of identifications by the total number of conclusive judgements. The estimates ranged from 100\% (i.e., all examiners correctly reported “identification”) down to 45\% (i.e., only 45\% of the examiners reported “identification”). This demonstrates substantial variability across individual firearms in terms of average examiner performance, at least in this study. 

An analogous result holds for the different-source test cases from \citep{mattijssen2020validity}, which were generated from 22 pairs of Glock pistols. Each pistol was fired once and each of the resulting 22 pairs of cartridge cases was evaluated by 77 examiners. For each pair of the 22 pairs, the true positive rate averaged over 77 examiners was estimated by dividing the number of exclusions by the total number of conclusive judgements. The estimates ranged from 100\% (i.e., all examiners correctly reported “exclusion”) down to 40\% (i.e., only 40\% of the examiners reported “exclusion”). This again demonstrates substantial variability across different pairs of the same firearm in average examiner performance.

\citet{chapnick2021results} use a study design that they reference from the \citet{montgomery2017design} textbook on experimental design, but they did not use the corresponding statistical method specified on the pages that they cited in this textbook for analyzing data when using that study design. That statistical method is designed to account for multiple sources of variation, such as the ones discussed above, but it was not used and instead an invalid method was used. 

To demonstrate the impact of taking variability across individual firearms (and pairs of individual firearms) into account when computing confidence intervals, consider \citep{monson2023accuracy}, which used 37 individual firearms for cartridge case comparisons. To appropriately compute confidence intervals, we advocate using the statistical model from \citet{gallas2009framework}, which is also used in the software referenced by the FDA. We advocate to combine this with an exact confidence interval method (since the estimated false positive probability is close to 0) that extends the commonly used Clopper-Pearson method to this setting. Setting aside all of the other flaws in Table 1, an exact 95\% confidence interval for the false positive error rate using this approach includes values greater or equal to 18\%. See Appendix B for a full description and proof. This is in contrast to the reported confidence interval (0.548\%, 1.57\%), which only has values going up to 1.57\%. The computation and justification for the 18\% depends only on the number of individual firearms used, the estimated false positive error rate from the study, and the study design. Since many of the other studies in Appendix A involve fewer individual firearms and similar estimated false positive rates, the approach in Appendix B can be used to show analogous results, i.e., that ignoring the impact of variability due to individual firearms can lead to much smaller confidence intervals than correctly taking this variability into account.

The remedy to the above problem, which occurs in all studies in Appendix A, is to apply one of the methods referenced above that is valid for analyzing data with multiple sources of variation.  Until that is done, it is unknown how much uncertainty to attach to the point estimates of error rates reported from these studies. Because of this, it is difficult or impossible to interpret the results from these studies (akin to learning the estimated percentages from an opinion poll without knowing the margin of error).

\subsection{Flaw E: Missing data}

Study validity requires that missing data, such as item- and unit-non-response, be handled by appropriate statistical methods. This was not done in any of the studies in Appendix A, with a few exceptions. Some of the smallest studies had no missing data reported, so flaw E does not apply. One of the methods used in the AMES II analysis, which computes examiner-specific error rates which it then averages across examiners, does implicitly address one part of missing data. However, that method is not sufficient for addressing missing data bias since it ignores potential factors that may impact both dropout and also estimator performance.

The importance of addressing missing data has been recognized at least since 2010 when the National Academies of Sciences published a seminal report on this topic for clinical studies \citep{nas2010prevention}.  In the peer-reviewed summary of their report, they state that ``Substantial instances of missing data are a serious problem that undermines the scientific credibility of causal conclusions from clinical trials.'' \citep[p.1355]{doi:10.1056/NEJMsr1203730}.  We agree, and missing data is a problem in all types of controlled experiments.  Black-box experiments involving firearms examination are no exception, and unfortunately the amount of missing data is substantial (where it is reported at all). For example, as described by \citet{khan2023shining}, the amount of missing data due to examiner dropout and non-response is greater than 30\% in \citet{monson2023accuracy}. At a minimum, every study should report the amount of missing data and conduct statistical analyses to address the additional bias and uncertainty introduced by it appropriately. This is not done in any of the firearms examination studies.

According to the National Academies of Science report on missing data in clinical trials  \citep[p.2]{nas2010prevention}:    \begin{quote}
Modern statistical analysis tools--such as maximum likelihood, multiple imputation, Bayesian methods, and methods based on generalized estimating equations--can reduce the potential bias arising from missing data by making principled use of auxiliary information available for non-respondents. The panel encourages increased use of these methods.
\end{quote}
These or other appropriate statistical methods can be used to reduce the potential bias caused by missing data in firearms examination studies. None of these methods  were used in any of the 28 firearms examination studies in Appendix A.

According to joint FDA and European Medicines Agency (EMA) guidances, the statistical analysis plan for every clinical trial should be written before the study starts \cite[p.27]{FDAEMAICHE92023} and should provide the ``Procedure for accounting for missing, unused, and spurious data.'' \citep[p.42]{FDAEMAgcp2018}. This applies to any validation study as well, regardless of whether it is a medical diagnostic test (for example) or a firearms examination black-box study. Failing to do an adequate statistical analysis to deal with missing data, as in the firearms examination studies (with the exception of some of the small studies that have no missing data), is invalid.

\section{Discussion} \label{sec:discussion}
We provided a statistical evaluation of 28 black-box validation studies, i.e., all such studies found in our literature search. We used the methodology of experimental design and guidance documents from the FDA about statistical standards in research studies. Our main finding is that methodological problems are pervasive and consequential, and thus the scientific validity of firearms examination has not been established.

Although current practice in firearms examination relies heavily on human judgements, researchers have been working on the development of automated, objective methods for forensic comparisons, as recommended by \cite{NAS2009, PCAST, kafadar2019}. Some black-box studies have already included  automated methods, e.g., \citet{mattijssen2020validity,Mattijssen2021, lawmorris2021}. Including automated methods as a ``participant'' in a black-box study is an opportunity to transition from human comparisons to algorithmic comparisons, or a hybrid combination of the two. A potential advantage of automated methods is increased transparency and replicability. In order to have  this advantage, however, automated algorithms need to be made open-source and all data in the corresponding validation studies (including that used to train the algorithms) need to be publicly available \citep{kafadar2019}. Currently, this is not the case for any of the black-box studies involving automated methods in our literature search. 

We evaluated the experimental design and statistical analysis methods used in studies that aim to assess the accuracy of firearms comparisons. We did not evaluate whether examiners in validation studies or real casework are conforming (i.e., adhering) to a clearly defined protocol or procedure. 
Conformance to such a protocol is needed in order to be able to generalize results from a study to a population of examiners (who use that protocol) in real casework as explained by \cite{mejia2020implementing,swofford2024talk}. 
The authors of a large black-box study \citep[p.4]{BALDWIN2023111739}, to their credit,  acknowledge the following about the set of possible conclusions that firearms examiners  can report:
\begin{quote}
The language used to define the AFTE Range of Conclusions is specific and concise, but our data suggest that interpretation and application of this system is not consistent across the profession. Rather, our analysis strongly suggests that, as currently applied, the categories of the current five-point AFTE Range of Conclusions scale cannot be said to have consistent, meaningful interpretations. \end{quote}
This lack of conformance to a well-defined, clearly interpretable, firearms examination protocol in practice is a major roadblock to any research program that aims to evaluate scientific validity of a forensic method. 
One needs to have both adequate conformance and adequate study design/analysis in order for a study to reliably assess scientific validity of a method \citep{swofford2024talk}.

We endeavored to provide useful recommendations for issues to consider and methods that can be used in the design and statistical analysis of future firearms examination validation studies. It also may be beneficial when planning future black-box studies to have an independent set of researchers and practitioners evaluate the proposed study design and statistical analysis plan before the study is started. This can help to find  and fix potential problems before it is too late, and such pre-reviews are standard in the conduct of medical research.

\section*{Disclosures and Acknowledgments}

This work was partially funded by the Center for Statistics and Applications in Forensic Evidence (CSAFE) through Cooperative Agreement 70NANB20H019 between NIST and Iowa State University, which includes activities carried out at Carnegie Mellon University, Duke University, University of California Irvine, University of Virginia, West Virginia University, University of Pennsylvania, Swarthmore College and University of Nebraska, Lincoln. Dr. Rosenblum was supported in this research by a Nexus Award from Johns Hopkins University. The opinions expressed herein are solely those of the authors and do not necessarily reflect the views of any institution or other person.  Some material in this document is reproduced, sometimes in a modified format, from Dr. Rosenblum’s report (declaration) in a criminal case where he is an expert witness for the defense \citep{homicide_case1}. Dr. Rosenblum is currently an expert witness in another case. His work as expert witness is through the consulting company Evolution Trial Design, Inc. of which  he is co-owner and president. Additional related work includes expert declarations co-authored by Drs. Susan Vanderplas, Kori Khan, Heike Hofmann, and Alicia Carriquiry (2022) that were filed in the aforementioned case, and an amicus curiae brief co-authored by Drs. Rosenblum, Vanderplas, and Cuellar, among others, for a case at the \cite{AbruquahAmicus}.

\bibliographystyle{Chicago}

\bibliography{references}

\newpage

\appendix
\section*{Appendix A: List of Black-Box Studies That We Reviewed}

Below we give references for black-box studies of firearms examination. Some studies are reported both in technical reports and in subsequently published papers, or are reported in multiple papers. In such cases, we only count each study once in determining how many studies the flaws in Table 1 apply to. That is why the total number of studies is 28 (and not 33, which is the length of the list).

The list includes two types of studies. The first type consists of black-box studies where examiners are presented with test cases of the following form: decide whether a single item (called the ``unknown'') was fired from the same gun as a set of reference items (called the ``knowns''). The items are either cartridge cases or bullets, and the reference items are fired from the same individual gun. 

Studies that use the other type of test cases, called ``set-based'' studies, not only have every flaw in Table 1 but also have more flaws.  In brief, ``set-based'' studies (e.g., where a set of items are all compared to each other, or where every item in one set needs to be matched to an item in another set) involve multiple comparisons in the same test question. The process of elimination (or similar logic) can be used to reduce the number of possibilities for the remaining comparisons after some comparisons have been decided on; this logical dependence can make the overall set of comparisons easier than if the comparisons had all been separate.

We included AFTE Journal articles but give a note of caution that there are potential issues with inadequate peer-review process during some years. 

\begin{enumerate}
\item Baldwin, D.P. Bajic, S.J., Morris, M., Zamzow, D. A study of false-positive and false-negative error rates in cartridge case comparisons. Technical report, AMES LAB IA, 2014.

\item Baldwin, D. P., and Stanley J. Bajic, Max D. Morris, Daniel S. Zamzow, A study of examiner accuracy in cartridge case comparisons. Part 1: Examiner error rates, Forensic Science International, Volume 349, 2023, \\
https://doi.org/10.1016/j.forsciint.2023.111733.

\item Baldwin, D. P., and Stanley J. Bajic, Max D. Morris, Daniel S. Zamzow, A study of examiner accuracy in cartridge case comparisons. Part 2: Examiner use of the AFTE range of conclusions, Forensic Science International, Vol. 349, 2023, 

\item Bajic, S.J., Chumbley, L.S., Morris, M., and Zamzow, D. Report: Validation study of the accuracy, repeatability, and reproducibility of firearm comparisons. Technical Report \# ISTR- 5220, Ames Laboratory-USDOE, 2020.

\item Best, B.A. \& Gardner, E.A. (2022). An assessment of the foundational validity of firearms identification using ten consecutively button-rifled barrels. AFTE Journal, 54(1), 28-37.

\item Brundage, D.J. (1998). The identification of consecutively rifled gun barrels. AFTE Journal, 30(3), 438-444.

\item Bunch, S.G., and Murphy, D.P. A Comprehensive Validity Study for the Forensic Examination of Cartridge Cases. AFTE J. 2003, 35(2), 201-203.

\item Cazes, M., \& Goudeau, J. (2013). Validation study of results from Hi-point consecutively manufactured slides. AFTE Journal, 45(2), 175-177.

\item Chapnick, C., Weller, T.J., Duez, P., Meschke, E., Marshall, J. and Lilien, R. (2021), Results of the 3D Virtual Comparison Microscopy Error Rate (VCMER) Study for firearm forensics. J Forensic Sci, 66: 557-570. \\ https://doi.org/10.1111/1556-4029.14602

\item DeFrance and Van Arsdale, M.D. Validation Study of Electrochemical Rifling. AFTE J. 2003. 35(1).

\item Duez, P., Weller, T., Brubaker, M., Hockensmith, R.E., II and Lilien, R. (2018), Development and Validation of a Virtual Examination Tool for Firearm Forensics. J Forensic Sci, 63: 1069-1084. https://doi.org/10.1111/1556-4029.13668

\item Fadul, T.G. (2011). An empirical study to evaluate the repeatability and uniqueness of striations/impressions imparted on consecutively manufactured Glock EBIS gun barrels. AFTE Journal, 43(1), 37-44.

\item Fadul, T.G., Hernandez, G.A., Stoiloff, S., \& Gulati, S. (2013). An empirical study to improve the scientific foundation of forensic firearm and tool mark identification utilizing 10 consecutively manufactured slides. AFTE Journal, 45(4), 376-391.

\item Fadul, T.G., Hernandez, G.A., Wilson, E., Stoiloff, S., \& Gulati, S. (2013) An Empirical Study To Improve The Scientific Foundation Of Forensic Firearm And Tool Mark Identification Utilizing Consecutively Manufactured Glock EBIS Barrels With The Same EBIS Pattern. Technical Report 244232 NCJRS.

\item Guyll M, Madon S, Yang Y, Burd KA, Wells G. Validity of forensic cartridge-case comparisons. Proceedings of the National Academy of Sciences. 2023 May 16;120(20):e2210428120.

\item Hamby, J.E., Brundage, D.J., \& Thorpe, J.W. (2009). The identification of bullets fired from 10 consecutively rifled 9mm Ruger pistol barrels: A research project involving 507 participants from 20 countries. AFTE Journal,41(2), 99-110.

\item Hamby, J.E., Brundage, D.J., Petraco, N.D.K. and Thorpe, J.W. (2019), A Worldwide Study of Bullets Fired From 10 Consecutively Rifled 9MM RUGER Pistol Barrels--Analysis of Examiner Error Rate. J Forensic Sci, 64: 551-557.\\ https://doi.org/10.1111/1556-4029.13916

\item Keisler, M.A. \& Hartman, S. \& Kilmon, A. \& Oberg, M. \& Templeton, M. (2018). Isolated pairs research study. AFTE Journal. 50. 56-58.

\item Knapp, J. and Garvin, A. (2012), Consecutively manufactured .25 auto F.I.E. barrels--A validation study. Presentation at AFTE 43rd Annual Training Seminar, Buffalo, NY. 

\item Law, E.F. and Morris, K.B. (2021), Evaluating firearm examiner conclusion variability using cartridge case reproductions. J Forensic Sci, 66: 1704-1720.\\ https://doi.org/10.1111/1556- 4029.14758

\item Lyons, D.J. (2009). The identification of consecutively manufactured extractors. AFTE Journal, 41(3), 246-256.

\item Mattijssen, EJAT, Witteman, CLM, Berger, Berger, CEH, Brand, N.W., Stoel, R.D. Validity and reliability of forensic firearm examiners, Forensic Science International, Volume 307, 2020, 110-112, https://doi.org/10.1016/j.forsciint.2019.110112.

\item Mattijssen, EJAT, Witteman, CLM, Berger, CEH, Zheng, XA, Sons, JA, Stoel, R.D., Firearm examination: Examiner judgments and computer-based comparisons. J Forensic Sci. 2021; 66: 96- 111. https://doi.org/10.1111/1556-4029.14557

\item Mayland, B., \& Tucker, C. (2012). Validation of obturation marks in consecutively reamed chambers. AFTE Journal, 44(2), 167-169.

\item Mikko, Don. An Empirical Study/Validation Test Pertaining to the Reproducibility of Toolmarks on 20,000 Bullets Fired Through M240 Machine Gun Barrels. AFTE Journal 45(3), 2013

\item Monson, K.L., Smith, E.D., Bajic, S.J. Planning, design and logistics of a decision analysis study: The FBI/AMES study involving forensic firearms examiners. Forensic Science International: Synergy, 4:100221, 2022.

\item Monson, K.L., Smith, E.D., and Peters, E.M. Accuracy of comparison decisions by forensic firearms examiners. Journal of forensic sciences, 68(1):86-100, 2023.

\item  Monson K.L., Smith E.D., Peters E.M.  Repeatability and reproducibility of comparison decisions by firearms examiners. Journal of Forensic Sciences 2023; 68: 1721–1740. https://doi.org/10.1111/1556-4029.15318 

\item Pauw-Vugts, P., Walters, A.,  Oren, L., Pfoser, L. (2013). FAID2009: Proficiency test and workshop. 45. 115-127. 

\item Smith E. Cartridge case and bullet comparison validation study with firearms submitted in casework. AFTE J. 2005; 37(4): 130-135.

\item Smith, T.P., Smith, A.G., \& Snipes, J.B. (2016). A validation study of bullet and cartridge case comparisons using samples representative of actual casework. Journal of Forensic Sciences, 61(4), 939-946.

\item Smith, J.A. (2021), Beretta barrel fired bullet validation study. J Forensic Sci, 66: 547-556. \url{https://doi.org/10.1111/1556-4029.14604}

\item Stroman, A. (2014). Empirically determined frequency of error in cartridge case examinations using a declared double-blind format. AFTE Journal, 46(2), 157-175.

\end{enumerate}

\section*{Appendix B: Confidence Interval Procedure for Black-Box Studies}
As described in Section~\ref{sec:confidenceIntervals}, the confidence intervals in all studies in our literature search are invalid since they fail to account for variation across firearms. 
We propose a direct extension of the Clopper-Pearson method that accounts for variation across both examiners and firearms. The method that we propose handles the resulting dependence structure (MRMC structure--see Section~\ref{sec:confidenceIntervals}) of the data, which applies to almost all of the black-box studies in Appendix A. 

We use the nonparametric statistical model from \cite{gallas2009framework}, which is also used in the software referenced by the \citep{FDA2007statistical} guidance on validation studies for diagnostic testing. This model applies to the MRMC data structure. We then define an exact confidence interval method (since the estimated false positive probability may be  close to 0) that extends the commonly used Clopper-Pearson method to this setting. Lastly, we describe 
 how this exact confidence interval method can be applied to data from the large black-box study of \cite{monson2023accuracy}, if we were to temporarily set aside all of the other flaws in Table 1. We show that the exact 95\% confidence interval for the false positive error rate using this approach includes values greater or equal to 18\%, which implies that it is much wider than the 95\% confidence interval reported in that study.

\textbf{Statistical Model:} 

The statistical model  defined by \cite[p. 2588]{gallas2009framework} is nonparametric except for independence assumptions given next. For simplicity, we focus on different source comparisons and the false positive error rate defined by \cite{monson2023accuracy} as the total number of identification responses divided by the total number of different source test cases\footnote{They exclude ``unsuitable" test cases, but we do not consider that here, for conciseness.} though the ideas can be applied to same source comparisons as well. When the \cite{gallas2009framework} model is applied to our context, it assumes that each individual firearm and examiner is an independent draw from independent populations, and that conditioned on an individual firearm (or pair of firearms for different source comparisons), the probability of a false positive response is independent across examiners, and that conditioned on each examiner the probability of a false positive is independent across individual firearms (or pairs of firearms for different sourse comparisons). The target of inference is the false positive rate averaged over the populations of examiners and individual firearms.

\textbf{Confidence Interval Procedure:}
We next define an exact 95\% confidence interval procedure for our problem (involving MRMC data) that is based on inverting hypothesis tests using a natural ordering of the sample space. This is precisely the approach used by, e.g., the Clopper-Pearson method for a sequence of $n$ independent, identically distributed data; there, the sample space is ordered by the number of 1's out of the total number of responses $n$. We use the analogous ordering applied to our problem. 
The sample space in our problem is defined to be all possible data sets that could result when using a given study design, where for simplicity each response is coded as 1 (``identification") or 0 (not ``identification"). 
 The ordering is by total number (denoted $m$)  of identification decisions across all examiners and test cases (denoted $n$).

The inputs to the confidence interval procedure are the following (using only different source comparisons, as explained above): the total number of ``identification" responses $m$, the total number of test cases $n$, and the study design (i.e., the plan for which test cases from each firearm or pair of firearms is evaluated by each examiner). Included in the study design is the number of individual firearms which we denote by $f$. In the study of \cite{monson2023accuracy}, for example, there are $f=37$ firearms involved for cartridge case comparisons (which we focus on below).
The output of the confidence interval procedure is a sub-interval of $[0,1]$ that roughly speaking, represents the range of plausible values of the studywide false identification rate.\footnote{We do not endorse this definition of the error rate, but since it was used by \cite{monson2023accuracy}, we use it as well in order to contrast our confidence interval procedure directly with the one from their paper.}

Consider any data generating distribution $P$ 
  on the sample space defined above that is in 
the aforementioned statistical model, i.e., for which the conditional independence relations defined by \cite{gallas2009framework} hold. Let $e(P)$ denote the parameter of interest, i.e., the population false identification rate, which is the probability of a false identification if a randomly selected  pair of firearms (having the same class characteristics) is examined by a randomly selected examiner. Lastly, let $P_0$ denote the true, unknown data generating distribution.

The confidence interval procedure for the false positive error rate $e(P)$ is based on inverting the two-sided, level 0.05 test of null hypothesis $H_0(p): e(P_0)=p$ (versus alternative hypothesis $e(P_0)\neq p$) for all  values of p in the unit interval [0,1]. The null hypothesis is composite, since it is defined to represent the class of all possible distributions $P$ on the sample space that are consistent with the aforementioned model and such that the false positive error rate $e(P)$  equals $p$. 

The following hypothesis testing procedure generalizes the Clopper-Pearson method to our statistical model. For any candidate value $p$ in [0,1], the null hypothesis $H_0(p)$ is rejected if for every distribution in $H_0(p)$, either the event $A$ (defined below) has probability at most 0.025 or the event $B$ (defined below) has probability at most 0.025. Event $A$ is the subset of the sample space where the number of identification responses (1's) is less than or equal to the observed number of identifications $m$; event $B$ is the subset of the sample space where the number of identification responses (1's) is greater than or equal to the observed number of identifications $m$. The confidence interval (technically, a confidence set) is defined to be the set of all $p$ in $[0,1]$ such that for at least one distribution in $H_0(p)$ the corresponding hypothesis test fails to reject. By construction, the hypothesis test has Type I error rate at most 0.05. There are other potential ways to compute a valid confidence interval than described above, but the above method is a direct generalization of the commonly used Clopper-Pearson method.

\textbf{Application to Study of \cite{monson2023accuracy}}:

We next prove that a valid, two-sided, 95\% confidence interval for the false positive error rate, based on \cite{monson2023accuracy}  study design and data from its different source cartridge case comparisons and using the above confidence interval procedure, includes values greater than 18\%. We prove this by constructing a data generating distribution $P$ in the statistical model with $e(P) =  0.1845$ for which the corresponding probability under $P$ of event A is at least 0.025 and the corresponding probability of event B is at least 0.025 and therefore $H_0(0.1845)$ cannot be rejected by the confidence interval procedure and so $p=0.1845$ is contained in the confidence interval.

Let $v = 0.0942$, which will be explained below. Define $\delta$ to be  the left endpoint of the Clopper-Pearson 95\% confidence interval for the false positive rate, e.g., $\delta = 0.006$ for the different-source cartridge case comparisons from  \cite{monson2023accuracy}. Let $m$ denote the number of false positives that occurred in the study out of $n$ total different-source comparisons. Therefore, by definition of the Clopper-Pearson confidence interval procedure, the probability that $m$ or more 1's occur in $n$ independent Bernoulli draws each with probability $\delta$ is 0.025.\footnote{This assumes that $m>0$. The case of $m=0$ is analogous.}

Define the following distribution $P$ on the sample space: 
\begin{quote}
Each individual firearm $j$ is represented as an independent, identically distributed draw $X_j$ from a Bernoulli($v$) distribution, i.e., it takes value 1 with probability $v$ and 0 otherwise. Consider any pair of firearms $i,j$, and any randomly drawn examiner $k$ who compares a single unknown bullet/cartridge case from one of these guns to one or more known bullets/cartridge cases from the other. If $X_i=X_j=0$, then the examiner makes a false positive error with probability $\delta$ based on an independent Bernoulli($\delta$) draw denoted $Y_{ijk}$; otherwise, if either $X_i=1$ or $X_j=1$, then the examiner makes a false positive error with probability 1. 
\end{quote}

Consider the corresponding probability that bullets/cartridge cases fired from two (different) randomly chosen firearms $i,j$ and examined by a randomly chosen examiner $k$ results in a false positive. This equals the probability that ($X_i$ or $X_j$ is 1) or ($X_i=X_j=0$ and $Y_{ijk}=1$), which by construction above is $p=2v-v^2+(1-v)^2 \delta = 0.1845$. Therefore, the above distribution is in $H_0(0.1845)$.

Recall that $m$ denotes the number of false positives that occurred in the study out of $n$ total different-source comparisons. The probability of $m$ or fewer false positives (across all $n$ different source comparisons in the study) under the aforementioned distribution $P$ is at least the product of  $(1-v)^{f}$ (for $f=37$) and  the probability that at most $m$ 1's occur in $n$ independent Bernoulli draws each with probability $\delta$, i.e., the probability of $m$ or fewer false positives is at least $(1-v)^{37} \times  0.975>0.025$.
We now explain how we selected the value of $v$ defined above. We had chosen the value of $v$ above so as to approximately maximize ($p=2v-v^2+(1-v)^2$) under the constraint that $P(A) \geq 0.025$. 
In fact, by our construction, the probability under $P$ of event $A$ above is at least 0.025 regardless of how many false positives $m$ are observed, due to the ordering defined in the Confidence Interval Procedure above. Also, the probability under $P$ of event $B$ is at least $1-(1-v)^{37} = 0.974$ (which exceeds 0.025) since event $B$ occurs whenever at least one firearm $j$ has $X_j=1$, due to the structure of the study design given in Table 4 on page 110 of the AMES II technical report \cite{amesii}. Since both events $A$ and $B$ have probability greater than 0.025 under the previously defined distribution, it follows that the hypothesis test in the Confidence Interval Procedure defined above fails to reject and therefore $p=0.1845$ is contained in the 95\% confidence interval for the false positive error rate.
\end{document}